\documentclass[onecollarge,natbib]{svjour2}
\bibpunct{[}{]}{,}{n}{}{,} 
\smartqed  
\usepackage{graphicx}
\journalname{Few-Body Systems (APFB2011)}
\begin{document}

\title{\boldmath
Penta-quark states with hidden charm and beauty
}


\author{Bing-Song Zou
}


\institute{B.~S.~Zou\at
              Institute of High Energy Physics and Theoretical Physics Center
              for Science Facilities, Chinese Academy of Sciences, Beijing
100049, China \\
              \email{zoubs@ihep.ac.cn}           
}

\date{Received: date / Accepted: date}

\maketitle

\begin{abstract}
More and more hadron states are found to be difficult to be
accommodated by the quenched quark models which describe baryons as
3-quark states and mesons as antiquark-quark states. Dragging out an
antiquark-quark pair from the gluon field in hadrons should be an
important excitation mechanism for hadron spectroscopy. Our recent
progress on the penta-quark states with hidden charm and beauty is
reviewed. 
\end{abstract}

\section{Quenched and unquenched quark models}
\label{intro}

In the classical quenched quark models, all established baryons are
ascribed into simple 3-quark (qqq) configurations~\cite{PDG}. The
classical quark models gave very good description of the mass
pattern and magnetic moments for the baryon SU(3) baryon $1/2^+$
octet and $3/2^+$ decuplet of spatial ground states. The excited
baryon states are described as excitation of individual constituent
quarks, similar to the cases for atomic and nuclear excitations. The
lowest spatial excited baryon is expected to be a ($uud$) $N^*$
state with one quark in orbital angular momentum $L=1$ state, and
hence should have negative parity. However, experimentally, the
lowest negative parity $N^*$ resonance is found to be $N^*(1535)$,
which is heavier than two other spatial excited baryons:
$\Lambda^*(1405)$ and $N^*(1440)$. This is the long-standing mass
reverse problem for the lowest spatial excited baryons.

In the simple 3q constituent quark models, it is also difficult to
understand the strange decay properties of the $N^*(1535)$, which
seems to couple strongly to the final states with strangeness.
Besides a large coupling to $N\eta$, a large value of
$g_{N^*(1535)K\Lambda}$ is deduced~\cite{liubc,gengls} by a
simultaneous fit to BES data on $J/\psi\to\bar pp\eta$,
$pK^-\bar\Lambda+c.c.$, and COSY data on $pp\to pK^+\Lambda$. There
is also evidence for large $g_{N^*(1535)N\eta^\prime}$ coupling from
$\gamma p \to p\eta^\prime$ reaction at CLAS~\cite{etap} and $pp\to
pp\eta^\prime$ reaction~\cite{caox}, and large $g_{N^*(1535)N\phi}$
coupling from $\pi^- p \to n\phi$, $pp\to pp\phi$ and $pn\to d\phi$
reactions~\cite{xiejj1,Doring,caox2009}.

On the other hand, unlike atomic and nuclear excitations, the
typical hadronic excitation energies are comparable with constituent
quark masses. Hence to drag out a $q\bar q$ pair from gluon field
could be a new excitation mechanism besides the conventional orbital
excitation of original constituent quarks. Then the mass reverse
problem and the strange decay properties of the $N^*(1535)$ can be
easily understood by considering 5-quark components in
them~\cite{liubc,Helminen,zhusl}. The $N^*(1535)$ could be the
lowest $L=1$ orbital excited $|uud>$ state with a large admixture of
$|[ud][us]\bar s>$ pentaquark component having $[ud]$, $[us]$ and
$\bar s$ in the ground state. The $N^*(1440)$ could be the lowest
radial excited $|uud>$ state with a large admixture of
$|[ud][ud]\bar d>$ pentaquark component having two $[ud]$ diquarks
in the relative P-wave. While the lowest $L=1$ orbital excited
$|uud>$ state should have a mass lower than the lowest radial
excited $|uud>$ state, the $|[ud][us]\bar s>$ pentaquark component
has a higher mass than $|[ud][ud]\bar d>$ pentaquark component. The
lighter $\Lambda^*(1405)1/2^-$ is also understandable in this
picture. Its main 5-quark configuration is $|[ud][us]\bar u>$ which
is lighter than the corresponding 5-quark configuration
$|[ud][us]\bar s>$ in the $N^*(1535)1/2^-$. The large mixture of the
$|[ud][us]\bar s>$ pentaquark component in the $N^*(1535)$ naturally
results in its large couplings to the $N\eta$, $N\eta^\prime$,
$N\phi$ and $K\Lambda$.

\begin{figure}
\centering
  \includegraphics{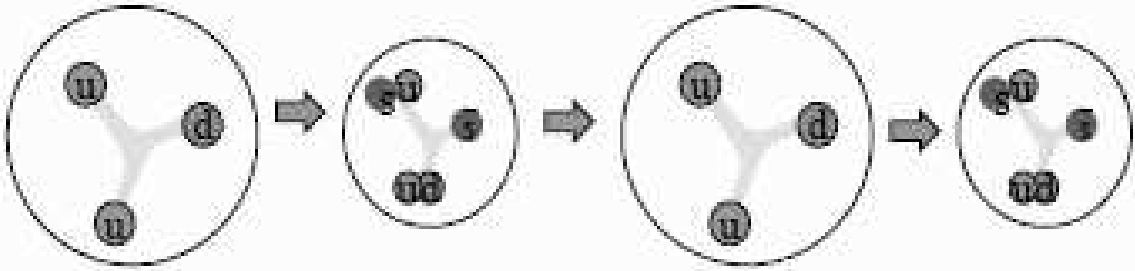}
\caption{The breathing mode of $qqq\leftrightarrow qqqq\bar q$ for
the lowest $1/2^-$ baryon octet.}
\label{fig:1}       
\end{figure}

A breathing mode of $qqq\leftrightarrow qqqq\bar q$ is
proposed~\cite{an2009,zouHypX} for the lowest $1/2^-$ baryon octet
as shown in Fig.\ref{fig:1}. Each baryon is a mixture of the
three-quark and five-quark components. The two components represent
two different states of the baryon. The $qqq$ state with $L=1$ and
higher kinetic energy has weaker potential; when quarks expand, a
$q\bar{q}$ pair is dragged out and results in a $qqqq\bar{q}$ state
with $L=0$ and stronger potential; the stronger potential leads
$qqqq\bar{q}$ state shrinking to a more compact state which then
makes the $\bar{q}$ to annihilate with a quark easily and transits
to the $qqq$ state with $L=1$ and more kinetic energy to expand;
this leads to constantly transitions between these two states. The
five-quark component has a smaller size than the three-quark
component and results in a much flatter $Q^2$-dependence for the
$\gamma^*N\to N^{*}(1535)$ transition where $\gamma^*qqq\to qqqq\bar
q$ plays a very important role~\cite{an2009,zouHypX}.

Besides the penta-quark configurations with the diquark correlation,
the penta-quark system may also be in the form of meson-baryon
states. The $N^*(1535)$, $\Lambda^*(1405)$ and some other baryon
resonances are proposed to be meson-baryon dynamically generated
states~\cite{Weise,or,Oset,meiss,Inoue,lutz,Hyodopk}.

Quenched $qqq$ quark models and unquenched $qqq\leftrightarrow
qqqq\bar q$ quark models give very different predictions for the
$1/2^-$ SU(3) nonet partners of the $N^*(1535)$ and
$\Lambda^*(1405)$. While quenched quark models~\cite{Capstick}
predict the $1/2^-$ $\Sigma^*$ and $\Xi^*$ to be around 1650 MeV and
1760 MeV, respectively, the unquenched quark
models~\cite{Helminen,zhusl,zouHypX} expect them to be around 1400
MeV and 1550 MeV, respectively, and meson-baryon dynamical
models~\cite{Oh,Kanchan,Ramos} predict them to be around 1450 MeV
and 1620 MeV, respectively.

Although various phenomenological models give distinguishable
predictions for the lowest $1/2^-$ $\Sigma^*$ and $\Xi^*$ states,
none of them are experimentally established. There is a $1/2^-$
$\Sigma^*(1620)$ listed as a 2-star resonance in the PDG
tables~\cite{PDG}. However, a recent analysis~\cite{gaopz} of the
new Crystal Ball data~\cite{prakhov09} on the $K^-p\to\pi^0\Lambda$
reaction with the $\Lambda$ polarization information indicates that
the $1/2^-$ $\Sigma^*(1620)$ is not needed by the data while the
$1/2^+$ $\Sigma^*(1635)$ is definitely needed. The data of
differential cross sections for this reaction without $\Lambda$
polarization information cannot distinguish the $1/2^-$ and $1/2^+$.
This may be the reason that some previous analyses claimed the
observation of the $1/2^-$ $\Sigma^*(1620)$. Instead some
re-analyses~\cite{wu_dulat1,wu_dulat2,gaopz2} of the $\pi\Lambda$
relevant data suggest that there may exist a $\Sigma^*(1/2^-)$
resonance around 1380 MeV, which supports the prediction of
unquenched quark models.

To pin down the nature of the lowest $1/2^-$ SU(3) baryon nonet, it
is crucial to find hyperon states of the lowest SU(3) $1/2^-$ nonet
and study their properties systematically. It would be very useful
to systematically analyze the available data on $K N\to K N$,
$\pi\Lambda$, $\pi\Sigma$ reactions and relevant $\psi$ decay
channels, including the new data from the Crystal Ball
Collaboration~\cite{prakhov09} and BESIII Collaboration.

\section{Prediction of superheavy $N^*$ and $\Lambda^*$ states with hidden charm and beauty
} \label{sec:1}

Although many $N^*$ and $\Lambda^*$ resonances were proposed to be
meson-baryon dynamically generated states or penta-quark states,
none of them can be clearly distinguished from qqq-model states due
to tunable ingredients and possible large mixing of various
configurations in these models. A possible solution to this problem
is to extend the penta-quark study to the hidden charm and hidden
beauty sectors. If the $N^*(1535)$ is the $\bar K\Sigma$ quasi-bound
state with hidden strangeness, then naturally by replacing $s\bar s$
by $c\bar c$ or $b\bar b$ one would expect super-heavy $N^*$ states
with hidden charm and hidden beauty just below $\bar D\Sigma_c$ and
$B\Sigma_b$ thresholds, respectively.

Following the Valencia approach of Ref.\cite{ramos} and extending it
to the hidden charm sector, the interaction between various charmed
mesons and charmed baryons were studied with the local hidden gauge
formalism in Refs.\cite{Wujj1,Wujj2}. Several meson-baryon
dynamically generated narrow $N^*$ and $\Lambda^*$ resonances with
hidden charm are predicted with mass around 4.3 GeV and width
smaller than 100 MeV. The S-wave $\Sigma_c\bar D$ and $\Lambda_c\bar
D$ states with isospin I=1/2 and spin S=1/2 were also dynamically
investigated within the framework of a chiral constituent quark
model by solving a resonating group method (RGM) equation by
W.L.Wang et al.~\cite{Wangwl}. They confirm that the interaction
between $\Sigma_c$ and $\bar D$ is attractive and results in a
$\Sigma_c\bar D$ bound state not far below threshold. The predicted
new resonances definitely cannot be accommodated by quark models
with three constituent quarks. Because these predicted states have
masses above $\eta_cN$ and $\eta_c\Lambda$ thresholds, they can be
looked for at the forthcoming PANDA/FAIR and JLab 12-GeV upgrade
experiments. This is an advantage for their experimental searches,
compared with those baryons with hidden charms below the $\eta_cN$
threshold proposed by other approaches~\cite{Gobbi,Hofmann}.

The same meson-baryon coupled channel unitary approach with the
local hidden gauge formalism was extended to the hidden beauty
sector in Ref.\cite{Wujj3}. Two $N^*_{b\bar b}$ states and four
$\Lambda^*_{b\bar b}$ states were predicted to be dynamically
generated. Because of the hidden $b\bar{b}$ components involved in
these states, the masses of these states are all above 11 GeV while
their widths are of only a few MeV, which should form part of the
heaviest island for the quite stable $N^*$ and $\Lambda^*$ baryons.
For the Valencia approach, the static limit is assumed for the
t-channel exchange of light vector mesons by neglecting momentum
dependent terms. In order to investigate the possible influence of
the momentum dependent terms, the conventional Schrodinger Equation
approach was also used to study possible bound states for the
$B\Sigma_b$ channel by keeping the momentum dependent terms in the
t-channel meson exchange potential. It was found that within the
reasonable model parameter range the two approaches give consistent
predictions about possible bound states. This gives some
justification of the simple Valencia approach although there could
be an uncertainty of 10 - 20 MeV for the binding energies.

Production cross sections of the predicted $N^*_{b\bar{b}}$
resonances in $pp$ and $ep$ collisions were estimated as a guide for
the possible experimental search at relevant facilities in the
future. For the $pp \to pp \eta_b$ reaction, the best center-of-mass
energy for observing the predicted $N^*_{b\bar{b}}$ is $13\sim 25$
GeV, where the production cross section is about 0.01 nb. For the
$e^-p \to e^-p \Upsilon$ reaction, when the center-of-mass energy is
larger than 14 GeV, the production cross section should be larger
than 0.1 nb. Nowadays, the luminosity for pp or ep collisions can
reach $10^{33}cm^{-2}s^{-1}$, this will produce more than 1000
events per day for the $N^*_{b\bar{b}}$ production. It is expected
that future facilities, such as proposed electron-ion collider
(EIC), may discover these very interesting super-heavy $N^*$ and
$\Lambda^*$ with hidden beauty.

Very recently, the observation of the iso-vector meson partners of
the predicted $N^*_{b\bar{b}}$, $Z_b(10610)$ and $Z_b(10650)$, were
reported by Belle Collaboration~\cite{Zb}. This gives us stronger
confidence on the existence of the super-heavy island for the
$N^*_{b\bar{b}}$ and $\Lambda^*_{b\bar{b}}$ resonances.

\begin{acknowledgements}
I thank C.~S.~An, P.~Z.~Gao, F.~Huang, R.~Molina, E.~Oset,
W.~L.~Wang, J.~J.~Wu, L.~Zhao, Z.~Y.~Zhang for collaboration works
reviewed here. This work is supported by the National Natural
Science Foundation of China (NSFC) under grants Nos. 10875133,
10821063, 11035006 and by the Chinese Academy of Sciences under
project No. KJCX2-EW-N01, and by the Ministry of Science and
Technology of China (2009CB825200).
\end{acknowledgements}



\end{document}